\documentclass[conference,a4paper]{IEEEtran}
\IEEEoverridecommandlockouts
\usepackage{cite}
\usepackage{amsmath,amssymb,amsfonts}
\usepackage[inline,shortlabels]{enumitem}
\usepackage{cleveref}
\usepackage{algorithmic}
\usepackage{graphicx}
\usepackage{textcomp}
\usepackage{xcolor}
\usepackage{siunitx}
\usepackage{threeparttable}
\usepackage{subcaption}
\usepackage{arydshln}
\usepackage{fancyhdr} 
\usepackage{multirow} 
\usepackage{amsmath} 
\usepackage{nomencl}

\IEEEoverridecommandlockouts 

\def\BibTeX{{\rm B\kern-.05em{\sc i\kern-.025em b}\kern-.08em
    T\kern-.1667em\lower.7ex\hbox{E}\kern-.125emX}}

\graphicspath{ {./} }
\crefname{equation}{}{}
\Crefname{equation}{Equation}{Equations}
\crefname{section}{Section}{Sections}
\Crefname{section}{Section}{Sections}
\crefname{figure}{Fig.}{Figs.}      
\Crefname{figure}{Figure}{Figures}  
\crefname{table}{Table}{Tables}
\Crefname{table}{Table}{Tables}
\newlist{E}{enumerate}{1}

\makenomenclature
\begin{document}

\title{Gaming Strategies in\\European Imbalance Settlement Mechanisms}

 \author{
\IEEEauthorblockN{Seyed Soroush Karimi Madahi}
\IEEEauthorblockA{\textit{IDLab, Ghent University -- imec} \\
Ghent, Belgium \\
seyedsoroush.karimimadahi@ugent.be}
\and
\IEEEauthorblockN{Kenneth Bruninx}
 \IEEEauthorblockA{\textit{Technology, Policy \& Management} \\
 TU Delft \\
 Delft, The Netherlands \\
 k.bruninx@tudelft.nl}
\and
\IEEEauthorblockN{Bert Claessens}
\IEEEauthorblockA{\textit{Beebop} \\
Belgium \\
bert@beebop.ai}
\and
\IEEEauthorblockN{Chris Develder}
 \IEEEauthorblockA{\textit{IDLab, Ghent University -- imec} \\
Ghent, Belgium \\
chris.develder@ugent.be}
}

\maketitle

\begin{abstract}
Transmission System Operators (TSOs) rely on balancing energy provided by Balancing Service Providers (BSPs) to maintain the supply-demand balance in real time. Balance Responsible Parties (BRPs) can simultaneously deviate from their day-ahead schedules in response to imbalance prices, e.g., by controlling flexible assets such as batteries. According to the European Electricity Balancing Guideline, these imbalance prices should incentivize  BRPs performing such implicit or passive balancing to aid the TSO in restoring the energy balance.
In this paper, we demonstrate that BRPs are unintentionally offered the opportunity to exploit gaming strategies in European imbalance settlement mechanisms. This is enabled by a disconnect between sub-quarter-hourly dynamics that determine the imbalance prices and the financial settlement on a quarter-hourly basis.
We illustrate this behavior in a case study of the imbalance settlement mechanisms in Belgium and the Netherlands. Our results reveal that, in both countries, BRPs can, in theory, exploit the imbalance mechanism by increasing the instantaneous system imbalance during minutes within the quarter-hour that determine the imbalance price while still contributing to restoring the system balance for the rest of the quarter-hour. 
\end{abstract}


\nomenclature[A0]{{\textbf{Sets}}}{}
\nomenclature[A1]{$\text{a/mFRR}^{+/-}$}{Set of upward/downward aFRR/mFRR bids}
\nomenclature[A2]{$Q$}{Set of minutes in a quarter-hour period}
\nomenclature[B00]{{\textbf{Decision Variables}}}{}
\nomenclature[B01]{$\lambda$}{Final imbalance price of ISP [{\texteuro}/MWh]}
\nomenclature[B02]{$\lambda^{+/-/\text{mid}}$}{Upward (+)/downward ($-$)/mid imbalance price of the quarter hour [{\texteuro}/MWh]}
\nomenclature[B03]{$\lambda^{\text{aFRR}^{+/-}}$}{Upward/downward aFRR price of the quarter hour [{\texteuro}/MWh]}
\nomenclature[B04]{$\lambda_t^\text{aFRR/mFRR}$}{aFRR/mFRR price at time $t$ [{\texteuro}/MWh]}
\nomenclature[B05]{$\textrm{SOC}_t$}{BESS state of charge at time $t$}
\nomenclature[B06]{$V_t^\text{aFRR/mFRR}$}{Total activated aFRR/mFRR volume at time $t$ [MW]}
\nomenclature[B07]{$b_t^{a/m^{+/-}}$}{Activated power of upward/downward regulation bid $a$ or $m$ at time $t$ [MW]}
\nomenclature[B08]{$p_t^\text{dis/cha}$}{BESS discharge/charge power at time $t$ [MW]}
\nomenclature[B09]{$z_t^\text{BESS}$}{Binary variable to prevent simultaneous BESS charging \& discharging (1 = charging)}
\nomenclature[B10]{$z_t^\text{mFRR}$}{Binary variable to show mFRR activation (1 = mFRR is activated)}
\nomenclature[C00]{{\textbf{Parameters}}}{}
\nomenclature[C01]{$\Delta t$}{Decision-making time resolution [h]}
\nomenclature[C02]{$\eta_\text{dis/cha}$}{BESS discharging/charging efficiency}
\nomenclature[C03]{$\mu^{a/m^{+/-}}$}{Activation price of upward/downward regulation bid $a$ or $m$ for the quarter hour [{\texteuro}/MWh]}
\nomenclature[C04]{$B^{r^{+/-}}$}{Maximum power of upward/downward regulation bid $r$ for the quarter hour [MW]}
\nomenclature[C05]{$E_\text{BESS}$}{Maximum capacity of BESS [MWh]}
\nomenclature[C06]{$E_\text{position}$}{BRP position for the quarter hour [MWh]}
\nomenclature[C07]{$P_\text{BESS}$}{Maximum power of BESS [MW]}
\nomenclature[C08]{$\text{SI}_t$}{System imbalance at time $t$ [MW]}
\nomenclature[C09]{$\underline{\textrm{SOC}} / \overline{\textrm{SOC}}$}{Minimum/Maximum BESS state of charge}
\nomenclature[C10]{$\overline{V}^{\text{aFRR/mFRR}^{+/-}}$}{Total upward/downward bid aFRR/mFRR volume for the quarter hour [MW]}
\printnomenclature

\section{Introduction}
The adoption of variable renewable energy sources (vRES) plays a crucial role in the energy transition. However, these vRES have introduced further variability and uncertainty to the grid operation due to their weather-dependent nature. This is complicating the task of balancing demand and supply for transmission system operators (TSOs), resulting in increased use and cost of balancing services throughout Europe~\cite{algarvio2024double}.

To aid TSOs in maintaining the balance, Balance Responsible Parties (BRPs), such as energy companies, are 
incentivized to balance consumption and generation within their portfolio~\cite{ENTSO}. 
Any real-time deviation from their commercial trade schedule is penalized based on the imbalance price~\cite{khodadadi2023sequential}. 
As imbalance prices were notoriously difficult to forecast and response times of conventional assets were limiting, imbalance prices were perceived as penalties~\cite{van2024probabilistic}. As a result, the imbalance settlement mechanism incentivized BRPs to balance their portfolios.


With the advent of fast response, flexible assets such as battery energy storage systems (BESS), better forecasting techniques and advanced decision support tools, market participants now have the opportunity to seek out-of-balance positions for financial gain~\cite{vermeulen2025}. BRPs can provide implicit balancing services to the TSO while securing profits from the imbalance settlement mechanism~\cite{smets2024participation}. Various studies have been conducted to demonstrate the profitability of implicit balancing for both BRPs and the benefits for TSOs~\cite{karimi2024control,smets2023strategic,madahi2024distributional}.
However, all these studies neglect the sub-quarter hourly dynamics in the imbalance settlement mechanism that determine the imbalance price (Section \ref{sec:imb_mech}). Pavarini et al.~\cite{pavirani2024predicting} showed that the publication of minute-based balancing data can sometimes significantly influence the final imbalance price, which may be attributed to BRPs' implicit balancing actions.


The ability of BRPs to adjust their positions within quarter hours using BESS in response to minute-based data, and its consequent impact on final imbalance prices, provokes an important follow-up question: can BRPs intentionally exacerbate imbalance prices to increase their profits? In this paper, we investigate whether the European Balancing Guidelines (EBGL) permits gaming strategies for BRPs. Although the EBGL harmonized the general principles of the imbalance settlement mechanism, the rules differ in some details across European countries. Hence, we selected two countries with clearly different imbalance settlement designs, namely Belgium (which uses a single pricing method) and the Netherlands (which uses a mixture of single and dual pricing methods).


The main contributions of this paper can be summarized as follows:
\begin{itemize}
\item To the best of our knowledge, this is the first paper that investigates gaming strategies arising from fast-response devices, such as batteries, in imbalance settlement mechanisms.
\item We study two different imbalance settlement mechanism designs to assess the impact of market design on BRPs' gaming strategies.
\item We consider sub-quarter hourly dynamics in imbalance settlement mechanisms to model imbalance prices more precisely.

\end{itemize}



\section{Imbalance Settlement Mechanism}\label{sec:imb_mech}
To keep the grid frequency within the desired range, TSOs correct system imbalances in real time by activating the appropriate frequency restoration reserve (FRR) volume provided by Balancing Service Providers (BSPs)~\cite{vermeulen2025}. TSOs charge unbalanced BRPs through imbalance prices at the end of each imbalance settlement period (ISP). The general goal of the imbalance settlement mechanism, as outlined in the EBGL, is to ensure that BRPs contribute efficiently to maintaining system balance~\cite{ENTSO}. To achieve this objective, the imbalance settlement mechanism must be designed to penalize deviations that worsen system imbalance while rewarding those that help restore it.

The EBGL aims to harmonize the imbalance settlement mechanism in Europe
, e.g., in terms of
the imbalance pricing methodology and ISP length (15 minutes). BSPs are remunerated for the provided balancing energy in each dispatch cycle, which can be as short as 4 seconds, based on a marginal pricing approach. The resulting price is the balancing energy price, which forms the input of the imbalance price calculation. For the imbalance price calculation, all TSOs shall implement single imbalance pricing, where both BRP shortage and BRP surplus are subject to the same price. However, TSOs have the freedom to choose how balancing energy prices are converted into imbalance prices: Volume-weighted average pricing, and marginal pricing are the two main approaches for this purpose. Furthermore, TSOs can apply dual pricing under special conditions, such as when both positive and negative FRR activations occur during the ISP. 

In this paper, we focus on the Belgian and Dutch imbalance settlement mechanisms. Although both markets follow the EBGL, they use different methods for calculating imbalance prices. The Belgian TSO (Elia) uses the single pricing methodology, with imbalance prices calculated based on the volume-weighted average price of activated FRR and the sign of the total system imbalance~\cite{elia2023brp}. In the Netherlands, the imbalance pricing is a mixture of single and dual pricing, where the imbalance price is based on the marginal price of activated FRR and the activation status of balancing energy, known as the regulation state~\cite{tennet}. If the Dutch TSO (TenneT) activates both FRR directions without monotonic balance deltas (i.e., activated upward FRR volume minus activated downward FRR volume) during the ISP, regulation state 2 is triggered, resulting in dual pricing. \Cref{table:imbalance price calculation} summarizes imbalance price calculation for Belgium and the Netherlands under different system conditions.

\renewcommand{\arraystretch}{1.3} 
\begin{table}[h!]
\centering
\caption{Imbalance price calculation in Belgium and the Netherlands under different system conditions}
\begin{threeparttable}
\begin{tabular}{ccrcc}
\hline
\multirow{2}{*}{\textbf{Country}} & \multicolumn{2}{c}{\multirow{2}{*}{\textbf{Grid Situation}}} & \multicolumn{2}{c}{\textbf{BRP Position}} \\ \cline{4-5}
 & \multicolumn{2}{c}{} & \textbf{Shortage} & \textbf{Surplus} \\ \hline
\multirow{2}{*}{Belgium} 
 & \multirow{2}{*}{\shortstack{Total System\\Imbalance}} & $> 0$ &  \multicolumn{2}{c}{$\lambda^{-}$} \\
 &  & $< 0$ &  \multicolumn{2}{c}{$\lambda^{+}$}  \\ \hline
\multirow{4}{*}{Netherlands} 
 & \multirow{4}{*}{\shortstack{Regulation\\State\tnote{*}}} & $0$ 
    &  \multicolumn{2}{c}{$\lambda^\text{mid}$} \\
 & & $+1$ 
    &  \multicolumn{2}{c}{$\lambda^{+}$} \\
 &  & $-1$ 
    &  \multicolumn{2}{c}{$\lambda^{-}$} \\
 &  & $2$ 
    & $\max(\lambda^\text{mid},\lambda^{+})$ & $\min(\lambda^\text{mid}, \lambda^{-})$ \\ \hline
\end{tabular}
\begin{tablenotes}
    \item[*] States are defined as:
    0~=~no FRR activation; 
    $+$1~=~upward FRR activation or both FRR directions activation with continuously increasing balance deltas; 
    $-$1~=~downward FRR activation or both FRR directions activation with continuously decreasing balance deltas; 
    2~=~both FRR directions activation without monotonic balance deltas.
\end{tablenotes}
\end{threeparttable}
\label{table:imbalance price calculation}
\end{table}

\section{Problem Formulation} \label{sec:meth}
\label{sec:problem formulation}
We study a BRP that controls its battery to maximize its profit from the imbalance settlement mechanism. If the imbalance settlement mechanism is properly designed, this control logic should provide an implicit balancing service that supports the TSO in restoring grid balance. The ISP is 15 minutes, but the battery power dispatch within each ISP is optimized on a minute-by-minute basis. The resolution used to calculate FRR prices is one minute.\footnote{Elia and TenneT calculate aFRR prices based on a 4-second optimization cycle. However, since they publish balancing data (system imbalance, aFRR/mFRR activations, etc.) every minute, a one-minute resolution is the finest granularity we could choose for this study.} As the BRP's actions affect grid balance and, as a result, the final imbalance price of the quarter hour, we formulate the problem as a bi-level optimization problem. The upper-level (BRP profit maximization) and lower-level (balancing market clearing) problems are explained in detail in~\cref{sec:upper level,sec:lower level}. We assume perfect foresight throughout our analysis to focus on how imbalance pricing strategies influence BRP's actions. 

\subsection{Upper Level -- BRP Profit}
\label{sec:upper level}
The BRP maximizes its profit for a single quarter hour, with fixed battery energy consumption for that quarter hour. 
We assume the BRP schedules its net position (total BESS energy consumption) for the quarter hour prior to the start of that quarter hour. Thus, under this assumption, we investigate the effect of minute-based decision-making on the final imbalance price and BRP profit. \Cref{1,2,3,4,5,6,7} show the upper-level optimization problem.
The objective function \cref{1} maximizes battery revenue in the imbalance settlement mechanism. \Cref{2,3} formulate the BESS state of charge (SoC) update and its corresponding SoC limitations. \Cref{4,5,6} guarantee that the (dis)charging energy for each minute does not exceed the maximum allowed energy and that the BESS does not charge and discharge simultaneously. \Cref{7} ensures that the BESS offers a predefined amount of balancing energy over the ISP ($E_\text{position}$).
\vspace{-0.5ex}
\begin{equation}
   \max_{p_t^\text{cha},p_t^\text{dis}} \>\> \lambda \> E_\text{position}
    \label{1}
\end{equation}
\vspace{-4ex}
\begin{align}
    &\text{Subject to:} \notag \\[-0.5ex]
    &\textrm{SOC}_{t+1} = \textrm{SOC}_t + \left(p_t^\text{cha} \eta_\text{cha} - \frac{p_t^\text{dis}}{\eta_\text{dis}} \right) \frac{\Delta t}{E_\text{BESS}} && \forall t \in Q \label{2} \\[-0.5ex]
    &\underline{\textrm{SOC}} \leq \textrm{SOC}_t \leq \overline{\textrm{SOC}} && \forall t \in Q \label{3} \\[-0.5ex]
    &0 \leq p_t^\text{cha} \leq z_t^\text{BESS} P_\text{BESS} && \forall t \in Q \label{4} \\[-0.5ex]
    &0 \leq p_t^\text{dis} \leq \left(1-z_t^\text{BESS}\right) P_\text{BESS} && \forall t \in Q \label{5} \\[-0.5ex]
    &z_t^\text{BESS} \in \{0,1\} && \forall t \in Q \label{6} \\[-0.5ex]
    &E_\text{position} = \sum_{t \in Q} \left( p_t^\text{dis} - p_t^\text{cha} \right) \Delta t \label{7}
\end{align}







\subsection{Lower Level -- Balancing Market Clearing}
\label{sec:lower level}
The lower-level problem models the balancing market clearing process, in which TSOs minimize the cost of balancing energy activation. We adopt a convex market model to formulate the market. Since Elia and TenneT use different approaches for imbalance price calculation, we formulate a separate lower-level problem for each country.

\textbf{Belgium:}
The lower-level problem in Belgium is given in \cref{8,9,10,11,12,13,14,15,16,17}. The objective is to minimize the balancing activation cost \cref{8}. \Cref{9,10,11} enforce the activation limitations of aFRR and mFRR, and prioritize aFRR activation over mFRR. The activation of the required balancing energy is guaranteed in~\cref{12,13,14,15}. The final imbalance price for the quarter hour is calculated according to~\cref{16,17}.
\begin{align}
   \min \sum_{t \in Q} \bigg( 
   &\sum_{a^+ \in \text{aFRR}^{+}} \mu^{a^+} b_t^{a^+}
   - \sum_{a^- \in \text{aFRR}^{-}} \mu^{a^-} b_t^{a^-} \notag \\
   +&\sum_{m^+ \in \text{mFRR}^{+}} \mu^{m^+} b_t^{m^+}
   - \sum_{m^- \in \text{mFRR}^{-}} \mu^{m^-} b_t^{m^-} 
   \bigg)
   \label{8}
\end{align}
\vspace{-6ex}

\begin{align}
    &\text{Subject to:} \notag \\[-0.5ex]
    &-\overline{V}^{\text{aFRR}^{-}} \leq V_t^\text{aFRR} \leq \overline{V}^{\text{aFRR}^{+}} && \hspace{-7ex} \forall t \in Q \label{9} \\[-0.5ex]
    &-\overline{V}^{\text{mFRR}^{-}} \!\! z_t^\text{mFRR} \leq V_t^\text{mFRR} \leq \overline{V}^{\text{mFRR}^{+}} \!\! z_t^\text{mFRR} && \hspace{-7ex} \forall t \in Q \label{10} \\[-0.5ex]
    &z_t^\text{mFRR} = \begin{cases}
        1 &: V_t^\text{aFRR} = \overline{V}^{\text{aFRR}^{+}} \text{or } -\overline{V}^{\text{aFRR}^{-}} \\
        0 &: \text{else}
    \end{cases} \label{11}  
\end{align}
\begin{align}
    &V_t^\text{aFRR} + V_t^\text{mFRR} = p_t^\text{cha} - p_t^\text{dis} - \text{SI}_t && \hspace{-7ex} \forall t \in Q
    \label{12} \\[0.5ex]
    &\sum_{a^+ \in \text{aFRR}^{+}} \! \! \! \! b_t^{a^+}
   - \! \! \! \! \sum_{a^- \in \text{aFRR}^{-}} \! \! \! \! b_t^{a^-} \!=\! V_t^\text{aFRR} \hspace{0.5ex}: \lambda_t^\text{aFRR} && \hspace{-7ex} \forall t \in Q \label{13} \\[0.5ex]
    &\sum_{m^+ \in \text{mFRR}^{+}} \! \! \! \! b_t^{m^+}
   - \! \! \! \! \! \sum_{m^- \in \text{mFRR}^{-}} \! \! \! \! b_t^{m^-} \!=\! V_t^\text{mFRR} \hspace{0.5ex}: \lambda_t^\text{mFRR} && \hspace{-7ex} \forall t \in Q \label{14} \\[0.5ex]
   &0 \leq b_t^{r^{+/-}} \leq B^{r^{+/-}} && \hspace{-7ex} \forall t \in Q \label{15} \\[-0.5ex]
    &\lambda^{\text{aFRR}^{+/-}} = \frac {\sum_{a^{+/-} \in \text{aFRR}^{+/-}} \mu^{a^{+/-}} b_t^{a^{+/-}}} {\sum_{a^{+/-} \in \text{aFRR}^{+/-}} b_t^{a^{+/-}}} \label{16} \\[-0.5ex]
    &\lambda = \begin{cases}
        \lambda^{\text{aFRR}^{+}} &: \text{SI}_t \leq 0 \text{ \& } z_t^\text{mFRR}=0 \\
        \max(\lambda^{\text{aFRR}^{+}}, \lambda_t^\text{mFRR})  &: \text{SI}_t \leq 0 \text{ \& } z_t^\text{mFRR}=1 \\
        \lambda^{\text{aFRR}^{-}} &: \text{SI}_t > 0 \text{ \& } z_t^\text{mFRR}=0 \\
        \min(\lambda^{\text{aFRR}^{-}}, \lambda_t^\text{mFRR}) &: \text{SI}_t > 0 \text{ \& } z_t^\text{mFRR}=1 \\
    \end{cases}
    \label{17}
\end{align}

\textbf{The Netherlands:}
The Dutch lower-level problem is formulated as follows, with~\cref{19,20,21,22} describing the final imbalance price calculation. 

\begin{equation}
    \min \> \> \sum_{t \in Q} \left( 
   \sum_{a^+ \in \text{aFRR}^{+}} \mu^{a^+} b_t^{a^+}
   - \sum_{a^- \in \text{aFRR}^{-}} \mu^{a^-} b_t^{a^-}\right)
    \label{18}
\end{equation}
\vspace{-5ex}





\begin{align}
    &\text{Subject to:} \notag \\[-0.5ex]
    &\text{\hspace{3cm} \cref{9,10,11,12,13,14,15}} \notag \\[0.5ex]
    &\lambda^{+} = \max_{t \in Q,\; b_t^{a^+} > 0} \lambda_t^{\text{aFRR}} \label{19} \\[0.5ex]
    &\lambda^{-} = \min_{t \in Q,\; b_t^{a^-} > 0} \lambda_t^{\text{aFRR}} \label{20} \\[0.5ex]
    &\lambda^\text{mid} =\frac{\min_{a^+ \in \text{aFRR}^{+}} \mu^{a^+} + \max_{a^- \in \text{aFRR}^{-}} \mu^{a^-}}{2}
    \label{21} \\[0.5ex]
    &\lambda = \begin{cases}
        \lambda^\text{mid} &: \text{state} = 0 \\
        \lambda^{+} &: \text{state} = +1 \\
        \lambda^{-} &: \text{state} = -1 \\
        \max (\lambda^{+},\lambda^\text{mid}) &: \text{state} = 2 \; \& \;
        E_\text{position}<0\\
        \min (\lambda^{-},\lambda^\text{mid}) &: \text{state} = 2 \; \& \;
        E_\text{position} \geq 0\\
    \end{cases}
    \label{22}
\end{align}


\subsection{Solution strategy}
We apply a piecewise linear approximation to transform \cref{16} into a linear equation. We cast the bi-level optimization problems as single-level problems by replacing the lower-level problems with their KKT conditions. The resulting mixed-integer nonlinear programs (MINLP) are solved using Gurobi.

\section{Gaming Strategies} \label{sec:results}
We used 2023 imbalance data from Belgium and the Netherlands. We consider a BRP with a 10MW/20MWh battery with 90\% round-trip efficiency ($\eta_\text{dis/cha}$). We identified two typical situations for each country in which BRPs can game the imbalance market to maximize revenue by optimally dispatching their battery power within the quarter hour without changing their net imbalance position for that period. 

We illustrate these strategies in selected ISPs. We compare an "optimized" dispatch strategy (following from our bilevel optimization problem) to a "uniform" dispatch. In the latter, the BESS provides the same amount of balancing energy over the ISP, but does not vary its dispatch.

\subsection{Belgium}
\begin{figure}[t]
    \centering
    \includegraphics[width=0.9\linewidth]{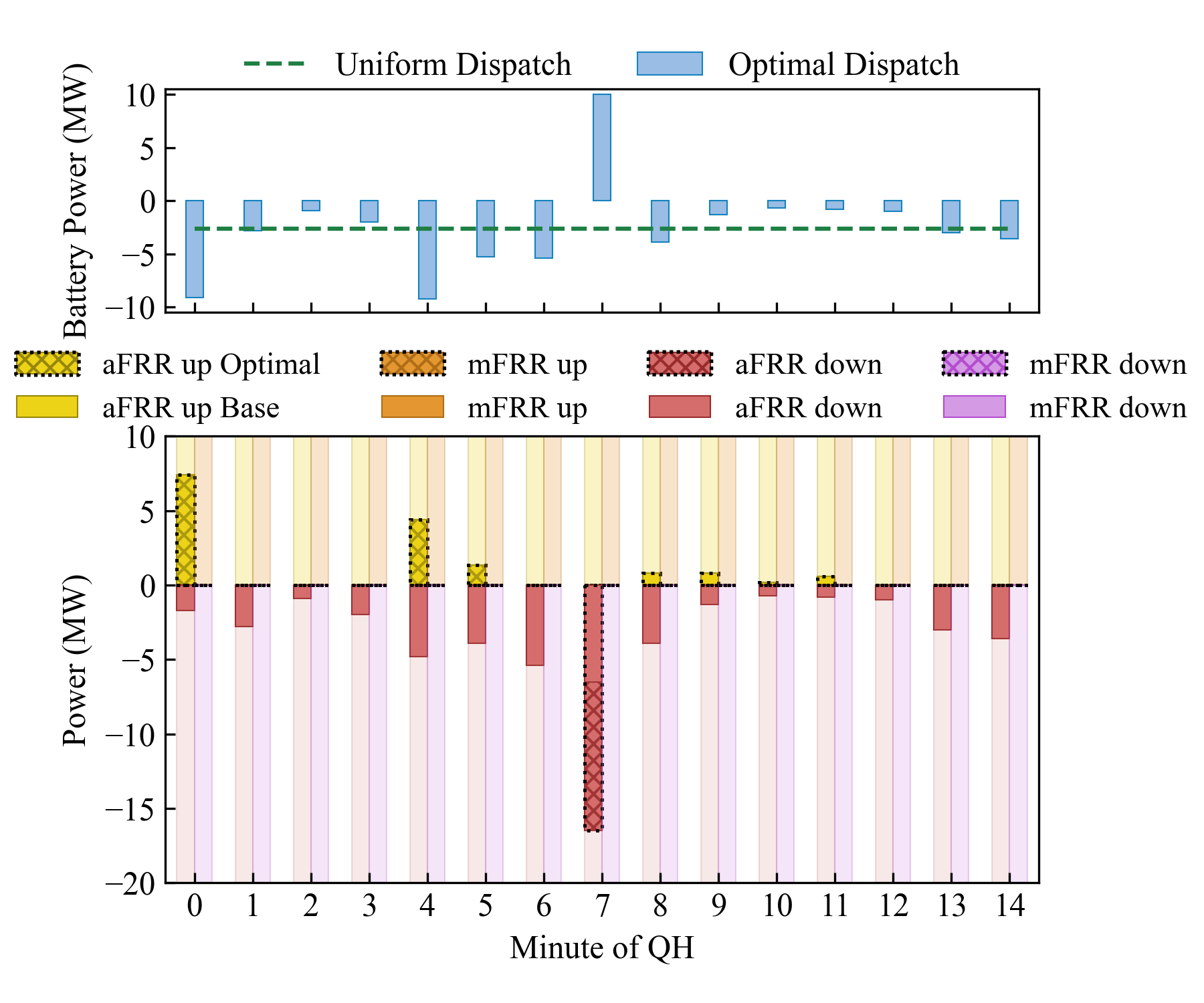}
    \caption{Optimization results for March 25, 2023, from 20:30 to 20:45 for the fixed average battery power of $-$2.6MW. (Top plot: Battery dispatch. Bottom plot: FRR activation)}
    \label{fig:within_QH_potential_2023_3_25_20_30}
\end{figure}

\textbf{Strategy 1 -- Activate Extreme FRR Bids:}
In this strategy, the BESS aggravates imbalance for 
at least
a minute within a quarter hour, triggering the activation of extreme FRR bids and consequently exacerbate the imbalance price for that quarter.
An example of this strategy is illustrated in~\cref{fig:within_QH_potential_2023_3_25_20_30}. The system imbalance for this quarter-hour is 2.66MW. The battery in the optimal dispatch case is consistently charged to support the grid—except for minute 7, when it is discharged at full power (note that positive battery power in the figures indicates battery discharge). The reason for this anomalous discharge is that, according to the bottom figure in~\cref{fig:within_QH_potential_2023_3_25_20_30}, the highest downward aFRR volume is activated at minute 7. Discharging during minute 7 increases the downward aFRR volume, triggering lower-priced bids and reducing the imbalance price from $-$84.75 \texteuro/MWh to $-$351.26 \texteuro/MWh. This increases the BRP's profit by 692.9\texteuro~ compared to the uniform dispatch.


\begin{figure}[t]
    \centering
    \includegraphics[width=0.9\linewidth]{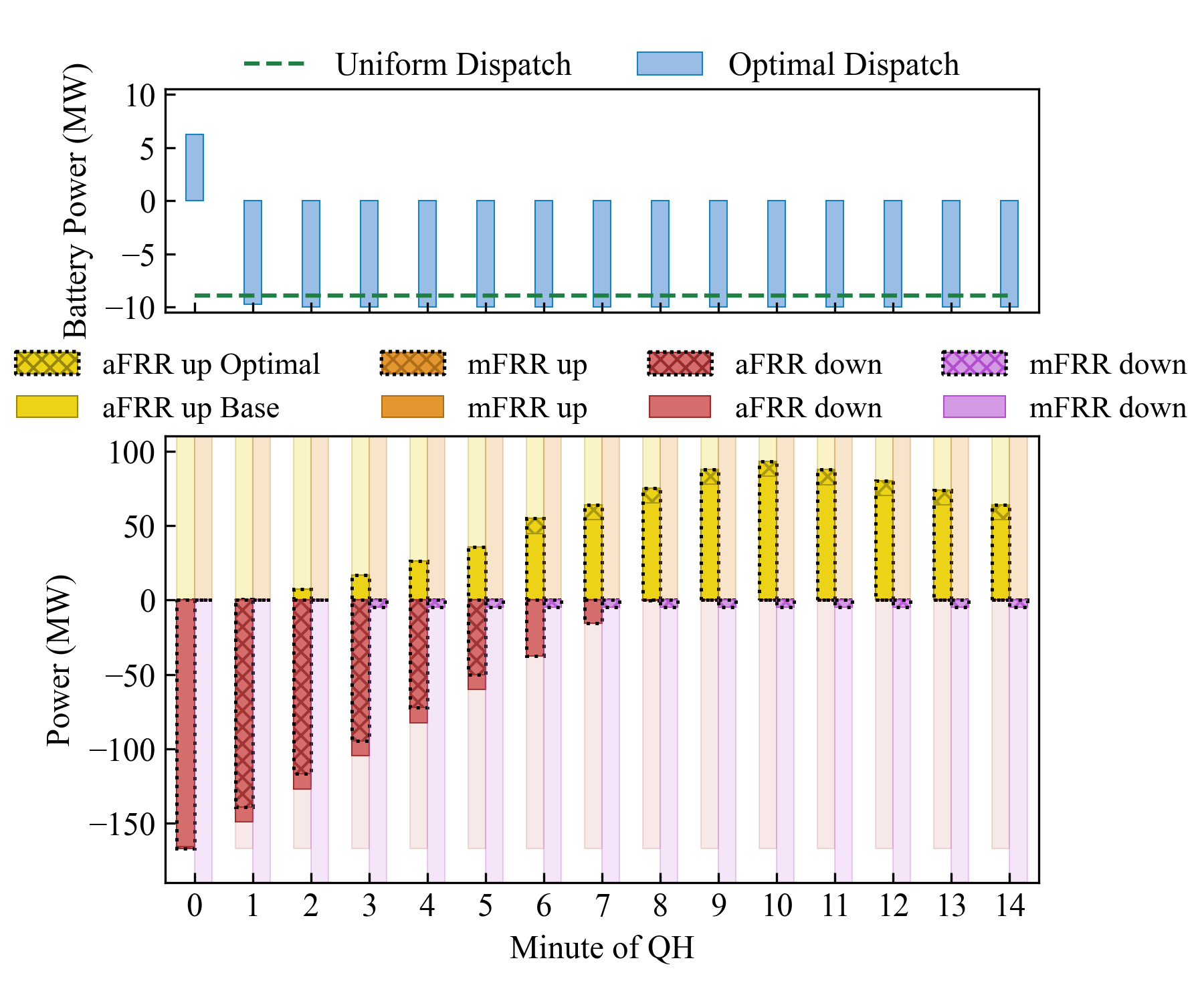}
    \caption{Optimization results for May 2, 2023, from 10:00 to 10:15 for the fixed average battery power of $-$8.9MW.}
    \label{fig:within_QH_potential_2023_5_2_10_0}
\end{figure}

\textbf{Strategy 2 -- Activate mFRR Bids:}
The amount of balancing energy provided under a uniform dispatch strategy may be limited because of the risk of reversing the system imbalance. On some occasions, such as in the quarter hour shown in \cref{fig:within_QH_potential_2023_5_2_10_0}, optimizing battery power within the ISP can avoid a change in the system imbalance direction by activating mFRR bids. The initial system imbalance for the mentioned quarter hour is 4.37MW. A uniform charging strategy at 8.9MW reverses the system imbalance direction, raises the price to 136.9\texteuro/MWh, and eventually causes the BRP to incur a loss of 304.6\texteuro. However, adopting the optimal power dispatch can raise the maximum average power to 8.9MW. In the first minute of the quarter hour, the battery is discharged to activate mFRR volume (\cref{fig:within_QH_potential_2023_5_2_10_0}). Since the triggered mFRR bid remains active until the end of the quarter hour, the battery can charge more without reversing the direction of the system imbalance which results in a profit of 195.3\texteuro~and the imbalance price of $-$87.8\texteuro/MWh. The mFRR activation in \cref{fig:within_QH_potential_2023_5_2_10_0} occurs at minute 3 because, for a direct mFRR activation, Elia sends the request 3 minutes before the mFRR activation.


\subsection{The Netherlands}
\begin{figure}[t]
    \centering
    \includegraphics[width=0.9\linewidth]{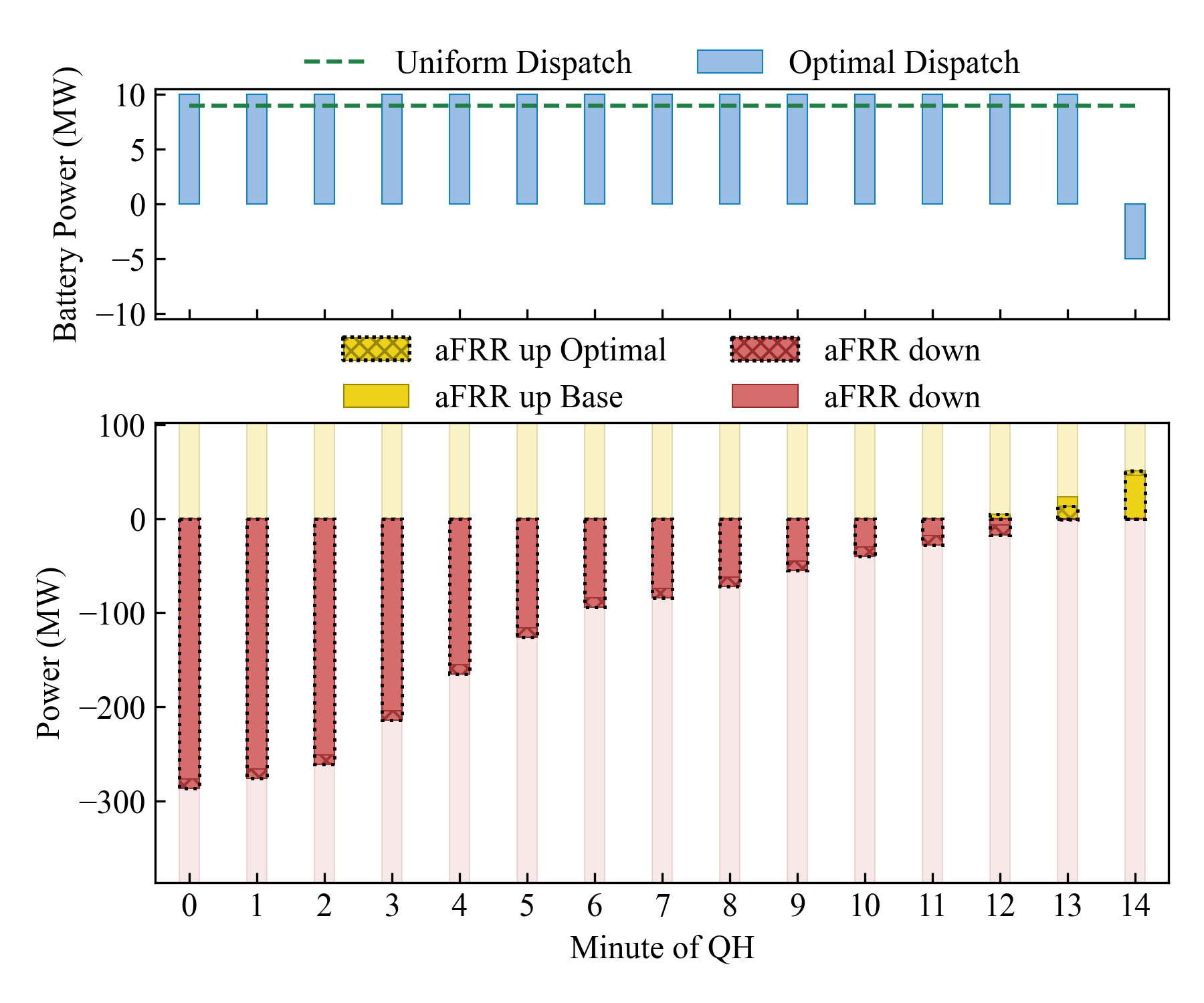}
    \caption{Optimization results for January 17, 2023, from 0:00 to 0:15 for the fixed average battery power of 9MW.}
    \label{fig:within_QH_potential_2023_1_17_0_0}
\end{figure}

\textbf{Strategy 1 -- Activate Extreme FRR Bids:}
The idea behind this strategy is similar to that of strategy 1 in the Belgian settlement. Since balance deltas continuously increase during the quarter hour illustrated in \cref{fig:within_QH_potential_2023_1_17_0_0}, the regulation state for this period is +1, meaning that the upward aFRR activations are the price setters. The battery is charged at minute 14 to trigger extreme aFRR bids, raising the imbalance price. Furthermore, throughout most of the period (except for minute 13), the battery aggravates the imbalance, which conflicts with the main objective of the imbalance settlement. 


\begin{figure}[t]
    \centering
    \includegraphics[width=0.9\linewidth]{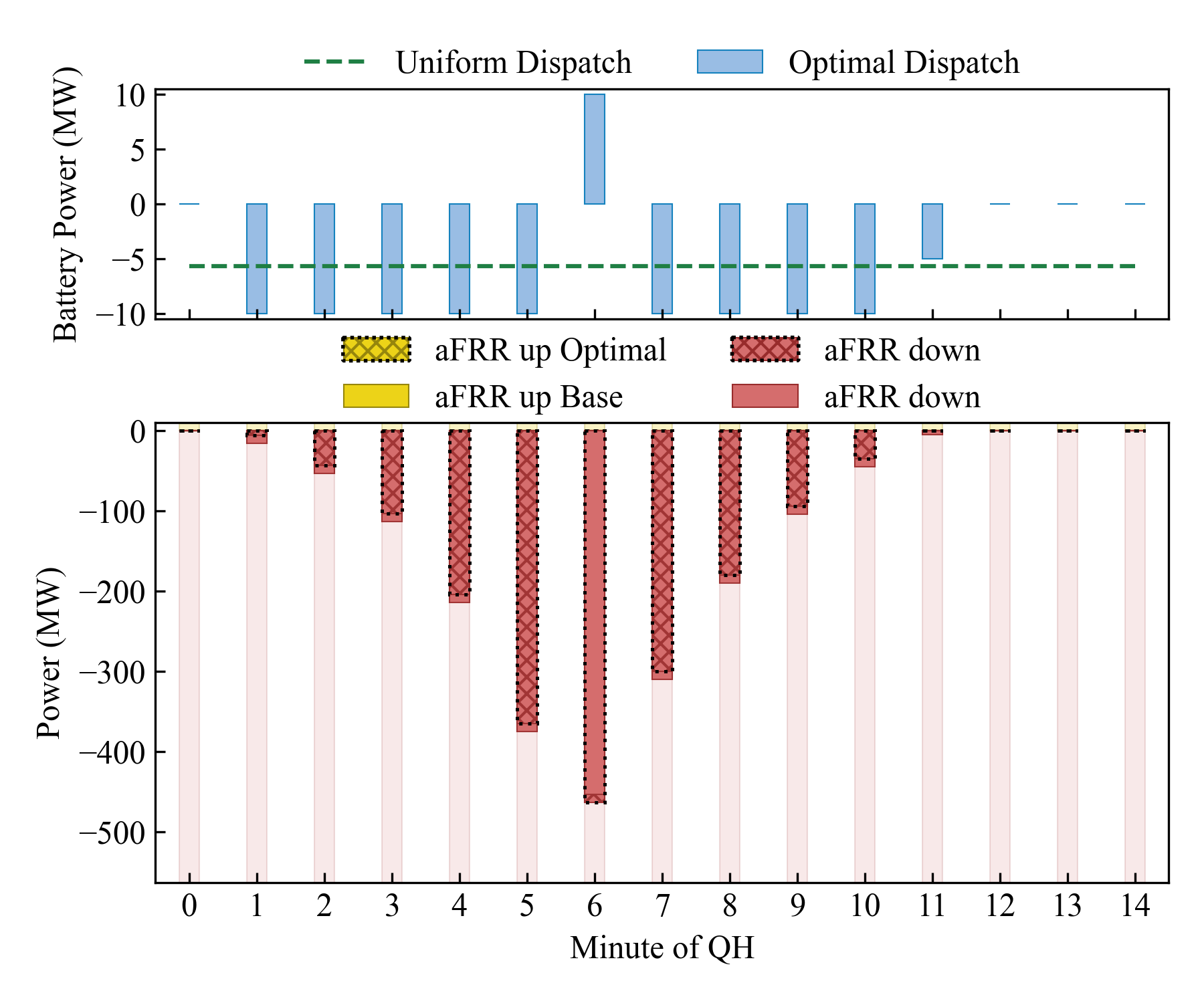}
    \caption{Optimization results for January 26, 2023, from 17:00 to 17:15 for the fixed average battery power of $-$5.7MW.}
    \label{fig:within_QH_potential_2023_1_26_17_0}
\end{figure}

\textbf{Strategy 2 -- Avoid Regulation State~2:}
The occurrence of regulation state~2 can sometimes be avoided by managing the battery power dispatch optimally. The ISP in~\cref{fig:within_QH_potential_2023_1_26_17_0} is an example of this situation. The regulation state in this ISP is -1 due to downward aFRR activations. A uniform discharge of the BESS at 5.7 MW causes upward activations of aFRR at several minutes, resulting in regulation state~2. In this case, the BRP is exposed to the upward price of 133.5\texteuro/MWh. However, the optimal power dispatch prevents the dual pricing mode by staying idle during minutes with close to zero system imbalance. Moreover, the BESS is discharged at minute 6 to activate extreme downward aFRR bids and further decrease the imbalance price. Thus, by optimally dispatching the battery power, the imbalance price becomes $-$255\texteuro/MWh, resulting in a profit gain of 550.39\texteuro~compared to the uniform scenario.

\section{Discussion \& Conclusion} \label{sec:conc}
All identified gaming strategies share a common feature: putting more pressure on system imbalance during price-defining moments in the ISP while restoring its aggregate balancing energy position in the rest of the ISP. As a result, they do not contribute to balancing the grid during critical minutes when the TSO urgently requires a balancing reaction, and they even exacerbate the stress on the system to make more profit. Studying the Dutch and Belgian imbalance settlement shows that activating extreme FRR bids is a common gaming strategy in both. The effect of this strategy on Dutch prices is more significant due to the adoption of the marginal pricing method. 

The example in~\cref{fig:within_QH_potential_2023_1_17_0_0} furthermore indicates that the current definition of the regulation state in the Dutch market sometimes does not fully support the main objective of the imbalance settlement design, as stated in the EBGL: BRPs may sometimes be rewarded for aggravating the imbalance and penalized for helping to restore the grid in the current Dutch market design. 


There are several potential solutions to eliminate these gaming strategies that warrant further research. The first one is to shorten the ISP to, for example, 5 minutes. BRPs that increase system imbalance during critical moments in the ISP see a higher impact on their final net position compared to the 15-minute ISP. Another possible solution is to modify the calculation of the imbalance position for BRPs. A counteracting imbalance factor can be added to the BRP position calculation, which determines how many minutes the BRP contributes to maintaining grid balance. 

\bibliographystyle{IEEEtran}
\bibliography{reference}

\end{document}